\documentclass{article}
\usepackage{spconf,amsmath,graphicx}
\usepackage{enumerate}
\usepackage{amssymb,amsfonts,graphicx}
\usepackage{cite}
\usepackage{algorithm}
\usepackage{algorithmicx}
\usepackage{algpseudocode}
\usepackage{amsthm}
\usepackage{mathrsfs}
\usepackage{nidanfloat}

\DeclareMathOperator*{\minimize}{minimize}

\DeclareMathOperator*{\subjectto}{subject~to}
\DeclareMathOperator*{\argmin}{argmin}

\DeclareMathOperator{\conv}{conv}

\DeclareMathOperator{\Tr}{Tr}
\def\inner<#1>{\langle #1 \rangle}

\title{Sparse time-frequency representation via atomic norm minimization}
\name{Tsubasa Kusano, Kohei Yatabe, Yasuhiro Oikawa}
\address{Department of Intermedia Art and Science, Waseda University, Tokyo, Japan}
\begin{document}
\ninept
\maketitle
\begin{abstract}
  Nonstationary signals are commonly analyzed and processed in the time-frequency (T-F) domain that is obtained by the discrete Gabor transform (DGT). The T-F representation obtained by DGT is spread due to windowing, which may degrade the performance of T-F domain analysis and processing.
  To obtain a well-localized T-F representation, sparsity-aware methods using $\ell_1$-norm have been studied.
  However, they need to discretize a continuous parameter onto a grid, which causes a model mismatch.
  In this paper, we propose a method of estimating a sparse T-F representation using atomic norm.
  The atomic norm enables sparse optimization without discretization of continuous parameters.   
  Numerical experiments show that the T-F representation obtained by the proposed method is sparser than the conventional methods.
\end{abstract}
\section{Introduction}
\label{sec:intro}

Nonstationary signals are commonly analyzed and processed in the time-frequency (T-F) domain.
For converting the time-domain signal into the T-F domain, the short-time Fourier transform (STFT) or the discrete Gabor transform (DGT) is usually utilized owing to its simplicity and well-understood structure \cite{Feichtinger1998,Grochenig2001}.
However, the T-F representation obtained by it is spread due to windowing of the analyzed signal.
This spread may affect the performance of T-F domain analysis and processing.

To achieve a well-localized T-F representation, many approaches have been proposed \cite{Auger1995,Daubechies2011,Meignen2012,Auger2013,Pfander2010,Gholami2013,Plumbley2010,Bayram2011,Balazs2013,Chen2014,Sejdic2018,Kowalski2018}.
Reassignment and synchrosqueezing methods aim to relocate the spread components into the original positions using phase derivative information \cite{Auger1995,Daubechies2011,Meignen2012,Auger2013}.
Their performances are affected by mixing of the components due to windowing \cite{Kusano2020}.
Sparsity-aware methods are powerful tools that are robust against such mixing of the components and noises \cite{Pfander2010,Gholami2013,Plumbley2010,Bayram2011,Balazs2013,Chen2014,Sejdic2018,Kowalski2018}.
Sparsity-aware methods aim to find a sparse solution of an underdetermined system.
However, the typical formulation based on $\ell_1$-norm minimization involves discretization of a continuous parameter onto a grid. It may degrade the performance due to a model mismatch between the signal and the predefined grid \cite{Chi2011}.

Recently, sparse optimization using atomic norm has been studied \cite{Candes2014,Chandrasekaran2012,Tang2013,Chi2020} and applied to many applications such as line spectrum estimation \cite{Yang2015,Li2016}, direction of arrival estimation \cite{Xenaki2015,Yang2017,Mahata2017}, and target localization in radar \cite{Heckel2016}.
Atomic norm does not require discretization of continuous parameters.
Thus, introducing the atomic norm into sparse T-F representation should obtain a better-localized T-F representation.

In this paper, we propose an estimation method of a sparse T-F representation.
In the proposed method, the estimation problem is formulated as atomic norm minimization under the condition that the analyzed time-domain signal can be reconstructed.
Numerical experiments confirmed that the proposed method provides a sparser T-F representation than the conventional methods.

Throughout this paper, ${\mathbb N}$, ${\mathbb R}$ and ${\mathbb C}$ denote the sets of all natural, real and complex numbers, respectively.
${\mathbf x}[n]$ is the $n$th element of a vector ${\mathbf x}$, and the ${\mathbf X}[m,n]$
is the $(m,n)$th element of a matrix ${\mathbf X}$.
$\overline{(\cdot)}$, ${(\cdot)}^\mathrm{T}$ and ${(\cdot)}^*$ represent the conjugate, the transpose and the conjugate transpose, respectively.
$\Tr({\mathbf X})$ is the trace of ${\mathbf X}$.
The inner product of vectors ${\mathbf x}, {\mathbf y}$ is defined as $\left\langle {\mathbf x}, {\mathbf y} \right\rangle = {\mathbf x}^\mathrm{T}\overline{\mathbf y}$.
The Frobenius inner product of two matrices ${\mathbf X}, {\mathbf Y}$ is defined as $\left\langle {\mathbf X}, {\mathbf Y} \right\rangle_{\mathrm F} = \Tr({\mathbf X}^\mathrm{T}\overline{\mathbf Y})$.
$\left\|\cdot \right\|_{\text F}$ denotes the Frobenius norm.
${\mathbf I}_L \in \mathbb{R}^{L\times L}$ is the identity matrix.
${\mathbf X} \succeq 0$ stands for ${\mathbf X}$ being positive semidefinite.

\section{Preliminaries}

\subsection{Gabor system and discrete Gabor transform (DGT)}

Let ${\mathbf g} \in \mathbb{R}^L$ denote a window.
A Gabor system is defined as \cite{Feichtinger1998,Grochenig2001}
\begin{equation}
  \mathcal{G}({\mathbf g}, a, M) = \left\{{\mathbf g}_{m,n}\right\}_{m = 0,\ldots, M-1,\, n=0,\ldots,N-1},
\end{equation}
where
\begin{equation}
  {\mathbf g}_{m,n}[l] = \mathrm{e}^{\mathrm{i}\frac{2\pi m (l-an)}{M}}{\mathbf g}[l-an],
\end{equation}
$a\in \mathbb{N}$ is the time-shifting width, and $M\in \mathbb{N}$ is the number of frequency channels.
DGT and the inverse DGT with respect to the Gabor system $\mathcal{G}({\mathbf g}, a, M)$ are defined by
\begin{equation}
  ({\mathbf G}_{\mathbf g}^*{\mathbf f})[m+nM] = \left\langle {\mathbf f}, {\mathbf g}_{m,n} \right\rangle,\qquad
  {\mathbf G}_{\mathbf g}{\mathbf c} = \sum_{m,n} {\mathbf c}[m+nM]{\mathbf g}_{m,n}, \nonumber
\end{equation}
where ${\mathbf c} \in {\mathbb C}^{MN}$ is a collection of the coefficients corresponding to a T-F representation.
If $\mathcal{G}({\mathbf g}, a, M)$ is a frame \cite{Feichtinger1998,Grochenig2001}, there exist a dual frame $\mathcal{G}({\mathbf h}, a, M)$ which satisfies
\begin{equation}
  {\mathbf f} = \sum_{m,n} \left\langle {\mathbf f}, {\mathbf g}_{m,n} \right\rangle{\mathbf h}_{m,n}.
\end{equation}
That is, a T-F representation ${\mathbf c}$ satisfying ${\mathbf f} = {\mathbf G}_{\mathbf g}{\mathbf c}$ can be obtained by DGT with a dual window ${\mathbf h}$ associated with ${\mathbf g}$.
A standard constuction of the dual window is the canonical dual window:
\begin{equation}
  \tilde{\mathbf h} = ({\mathbf G}_{\mathbf g}{\mathbf G}_{\mathbf g}^*)^{-1}{\mathbf g}.
  \label{eq:canonicalDual}
\end{equation}



\subsection{Sparse T-F representation using $\ell_1$-norm}

A T-F representation obtained by DGT ${\mathbf G}_{\mathbf h}^*$ is spread due to multiplication of a dual window ${\mathbf h}$.
If ${\mathcal G}({\mathbf g}, a, M)$ is a frame, the T-F representation ${\mathbf c}$ is a redundant representation of a time-domain signal ${\mathbf f}$, i.e., the T-F representation ${\mathbf c}$ satisfying
\begin{equation}
  {\mathbf f} = {\mathbf G}_{\mathbf g}{\mathbf c}
  \label{eq:perfect_reconstruction}
\end{equation}
is not unique.
The direct formulation for finding the sparsest solution of this underdetermined system is to minimize the number of non-zero coefficients, called the $\ell_0$-norm.
Unfortunately, this problem is usually an intractable combinatorial optimization problem, and its solution is sensitive to noise.
Instead of the $\ell_0$-norm, the $\ell_1$-norm has been widely used as a cost function for promoting sparsity \cite{Pfander2010,Gholami2013,Plumbley2010,Bayram2011,Balazs2013,Chen2014,Sejdic2018,Kowalski2018}.
It is formulated as
\begin{equation}
  \minimize_{\mathbf c} \quad  \left\| {\mathbf c} \right\|_1 \quad  \subjectto \quad  {\mathbf f} = {\mathbf G}_{\mathbf g} {\mathbf c},
  \label{eq:l1_norm}
\end{equation}
where $\left\|\cdot \right\|_1$ is the $\ell_1$-norm, defined as $\left\| {\mathbf c} \right\|_1 = \sum_{j=0}^{NM-1}\left| {\mathbf c}[j]\right|$.
This problem is convex, which can be solved by convex optimization algorithms.
Nonetheless, ${\mathcal G}({\mathbf g}, a, M)$ contains windowed sinusoids whose frequency is discretized onto the grid $\left\{m/M \right\}_{m=0, \ldots, M-1}$.
Eq.~\eqref{eq:l1_norm} may provide a poor result when the signal ${\mathbf f}$ has a component whose frequency is not included in the grid.

\subsection{Line spectrum estimation using atomic norm}

\label{sec:lineSpectrumEstimation}

To avoid the effects of grid mismatch, a method using atomic norm has been studied as a gridless sparse optimization method \cite{Candes2014,Chandrasekaran2012,Tang2013,Chi2020,Yang2015,Li2016,Xenaki2015,Yang2017,Mahata2017,Heckel2016}.
Here, line spectrum estimation is applied only to the $n$th windowed signal.
A windowed signal at time index $n$ is denoted as ${\mathbf f}_n = {\mathbf W}_n{\mathbf f}$, where ${\mathbf W}_n \in \mathbb{R}^{L\times L}$ is a diagonal matrix whose diagonal elements are given by ${\mathbf W}_n[l, l] = {\mathbf g}[l-an]$.
We assume that the $n$th windowed signal ${\mathbf f}_n$ can be expressed as a sum of complex sinusoids,
\begin{equation}
  {\mathbf f}_n  = {\mathbf W}_n\sum_k c_{n,k} {\mathbf a}_{n,k},\qquad {\mathbf a}_{n,k} \in {\mathcal A},
\end{equation}
where ${\mathcal A}$ is a collection of complex sinusoids
\begin{equation}
  \mathcal{A} = \left\{ {\mathbf a}\in\mathbb{C}^L \,\middle|\, {\mathbf a}[l] = \mathrm{e}^{\mathrm{i}2\pi \omega l}, \omega \in [0 , 1)\right\}.
  \label{eq:atom_freq}
\end{equation}
The atomic norm is used to express the $n$th windowed signal ${\mathbf f}_n$ with a few coefficient $c_{n,k}$.
Let us denote ${\mathbf x}_n = \sum_k c_{n,k} {\mathbf a}_{n,k}$, then the atomic norm of ${\mathbf x}_n$ associated with a set of atoms ${\mathcal A}$ is given by \cite{Chandrasekaran2012}
\begin{align}
  \left\| {\mathbf x}_n \right\|_\mathcal{A} &= \inf\left\{\nu_n \ge 0 \,\middle|\, {\mathbf x}_n \in \nu_n \conv({\mathcal A}) \right\}, \label{eq:atomic_norm} \\
  &= \inf\left\{ \sum_k |c_{n,k}| \,\middle|\,{\mathbf x}_n = \sum_k c_{n,k} {\mathbf a}_{n,k},\,\,\, {\mathbf a}_{n,k} \in \mathcal{A}\right\}, \nonumber
\end{align}
where $\conv({\mathcal A})$ is the convex hull of ${\mathcal A}$.
It corresponds to the infimum of the $\ell_1$-norm of coefficients when ${\mathbf x}_n$ is represented by a linear combination of elements in ${\mathcal A}$.
That is, the atomic norm can be interpreted as an extension of the $\ell_1$-norm to the continuous parameter $\omega \in [0,1)$.
The atomic norm in Eq.~\eqref{eq:atomic_norm} is characterized by the following optimization problem \cite{Tang2013}:
\begin{align}
  \left\| {\mathbf x}_n \right\|_\mathcal{A} = & \min_{{\mathbf u}_n, \nu_n} \quad \frac{1}{2L}\Tr\left(T({\mathbf u}_n)\right) + \frac{1}{2} \nu_n \nonumber \\
  &\subjectto \quad  \left[\begin{array}{cc}
    T({\mathbf u}_n) & {\mathbf x}_n \\
    {\mathbf x}_n^* & \nu_n
  \end{array}\right] \succeq 0,
  \label{eq:atomic_norm_SDP}
\end{align}
where $T: {\mathbb C}^L \to {\mathbb C}^{L\times L}$ is the Hermitian Toeplitz operator:
\begin{equation}
  T({\mathbf u}) =
  \left[\begin{array}{cccc}
    {\mathbf u}[0] & \overline{{\mathbf u}[1]} & \cdots & \overline{{\mathbf u}[L-1]} \\
    {\mathbf u}[1] & {\mathbf u}[0] & \ddots & \vdots \\
    \vdots & \ddots & \ddots & \overline{{\mathbf u}[1]} \\
    {\mathbf u}[L-1] & \cdots & {\mathbf u}[1] & {\mathbf u}[0]
  \end{array}\right].
\end{equation}
If the Hermitian Toeplitz matrix $T({\mathbf u}_n)$ is positive semidefinite and singular, it can be uniquely decomposed as \cite{Caratheodory1911}
\begin{equation}
  T({\mathbf u}_n) = \sum_{k=0}^{K-1} \left| c_{n,k} \right| {\mathbf a}_{n,k}{\mathbf a}^*_{n,k},\qquad {\mathbf a}_{n,k} \in \mathcal{A},
  \label{eq:Vandermonde_decomposition}
\end{equation}
where $K$ corresponds to the rank of $T({\mathbf u}_n)$.
${\mathbf a}_{n,k}$ in Eq.~\eqref{eq:Vandermonde_decomposition} can be obtained by Prony's method \cite{RichedeProny1795}, the matrix pencil method \cite{Hua1990}, or other linear prediction methods \cite{Tufts1982}.
Then, coefficients $c_{n,k}$ can be obtained by solving the linear equation
\begin{equation}
  \left[{\mathbf a}_{n,0},\ldots, {\mathbf a}_{n,K-1} \right]{\mathbf c}_n = {\mathbf x_n},
\end{equation}
where ${\mathbf c}_n = \left[c_{n,0},\ldots, c_{n,K-1} \right]^\mathrm{T}$.
Therefore, estimating sinusoids in the windowed signal ${\mathbf f}_n$ using the atomic norm is formulated as
\begin{equation}
  \minimize_{\mathbf{x}_n}\quad \left\|{\mathbf x}_n \right\|_\mathcal{A} \quad \subjectto \quad {\mathbf f}_n = {\mathbf W}_n{\mathbf x}_n.
  \label{eq:ANM_BP}
\end{equation}
Substituting Eq.~\eqref{eq:atomic_norm_SDP} into Eq.~\eqref{eq:ANM_BP}, Eq.~\eqref{eq:ANM_BP} can be rewritten as the following semidefinite programming:
\begin{align}
  \minimize_{\mathbf{x}_n,\mathbf{u}_n, \nu_n}& \quad \frac{1}{2L}\Tr(T({\mathbf u}_n)) + \frac{1}{2} \nu_n \nonumber \\
  \subjectto& \quad \left[\begin{array}{cc}
    T({\mathbf u}_n) & {\mathbf x}_n \\
    {\mathbf x}_n^* & \nu_n
  \end{array}\right] \succeq 0,\quad {\mathbf f}_n = {\mathbf W}_n{\mathbf x}_n.
  \label{eq:ANM_BP_SDP}
\end{align}

While the eigenvalue decomposition of the $(L+1)\times (L+1)$ matrix needs to be iterated when solving Eq.~\eqref{eq:ANM_BP_SDP}, it can be reduced to the eigenvalue decomposition of a $(J+1)\times (J+1)$ matrix if the window function is supported on $[0, J-1]$.

\section{Proposed method}

\begin{figure}[t]
	\centering
	\includegraphics[width=0.96\columnwidth]{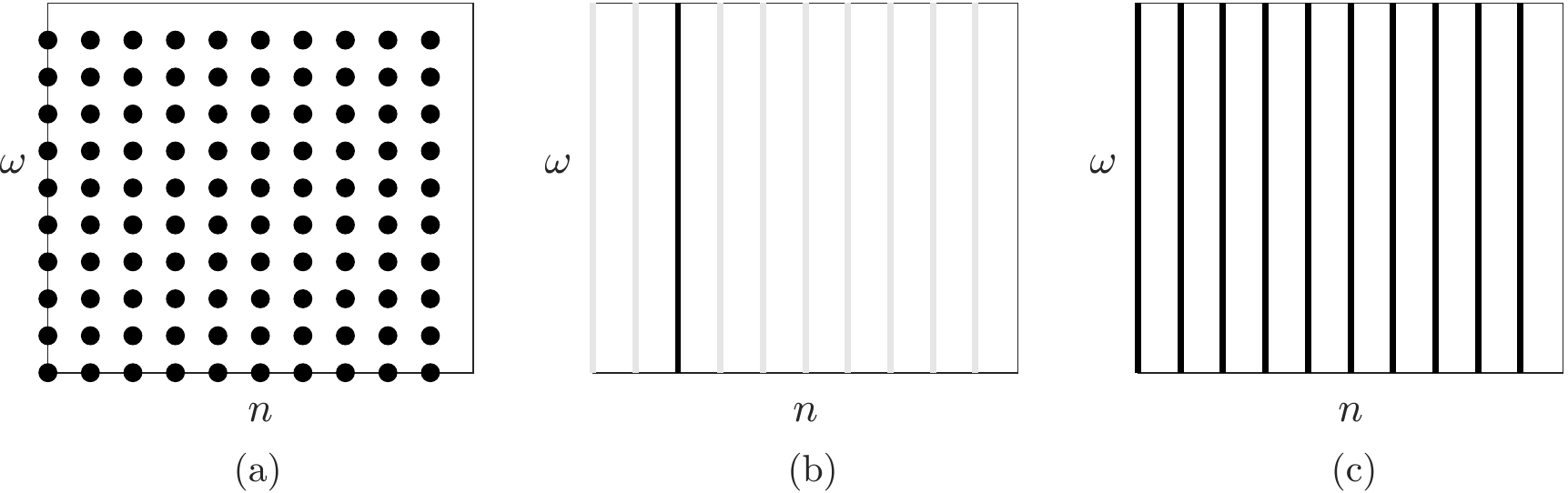}
      \vspace{-10pt}
	\caption{Estimating T-F representations using (a) $\ell_1$-norm, (b) the line spectrum estimation by atomic norm, and (c) the proposed method.}
	\label{fig:dictionaries}
\end{figure}

The method of estimating the T-F representation based on the $\ell_1$-norm suffers from degradation of performance due to discretization onto the grid as shown in Fig.~\ref{fig:dictionaries}~(a).
On the other hand, performing the line spectrum estimation using the atomic norm for each time index, a sparse T-F representation can be obtained without a grid in frequency direction.
However, Eq.~\eqref{eq:ANM_BP_SDP} estimates sinusoids at each time index $n$ independently as Fig.~\ref{fig:dictionaries}~(b), which does not efficiently take advantage of the sparsity of T-F representation (see Fig.~\ref{fig:Fig1}).

In this paper, we propose a estimation method of a sparse T-F representation using the atomic norm to avoid the effect of grid mismatch.
In our formulation, the atomic norm is minimized under the constraint of the perfect reconstruction of the entire signal.
It can be interpreted as an extension of Eq.~\eqref{eq:l1_norm} with an infinite number of frequency channels as Fig.~\ref{fig:dictionaries}~(c).

\subsection{Sparse T-F representation using atomic norm}

Reconstruction of the $n$th windowed signal ${\mathbf f}_n = {\mathbf W}_n{\mathbf f}_n$ is considered in Eq.~\eqref{eq:ANM_BP_SDP}.
In contrast, reconstructing the entire signal ${\mathbf f}$ by windowing and summing ${\mathbf x}_n$ is considered in the proposed method, which is written as
\begin{equation}
  {\mathbf A}_{\mathbf g}{\mathbf x} = \sum_{n=0}^{N-1} {\mathbf W}_n {\mathbf x}_n,
  \label{eq:perfect_reconstruction_ANM}
\end{equation}
where ${\mathbf x} = \left[{\mathbf x}_0^\mathrm{T}, {\mathbf x}_1^\mathrm{T}, \ldots, {\mathbf x}_{N-1}^\mathrm{T}\right]^\mathrm{T}$.
As a cost function, a sum of atomic norms for time index $n$ is chosen because it corresponds to the grid-less version of the cost function in Eq.~\eqref{eq:l1_norm},
\begin{equation}
  \sum_{n=0}^{N-1} \left\| {\mathbf x}_n \right\|_\mathcal{A} = \inf\left\{ \sum_{n,k} |c_{n,k}| \,\middle|\,{\mathbf x}_n = \sum_k c_{n,k} {\mathbf a}_{n,k},\,\,\, {\mathbf a}_{n,k} \in \mathcal{A}\right\} . \nonumber
\end{equation}
Thus, the estimation of a sparse T-F representation is formulated as
\begin{equation}
  \minimize_{{\mathbf x}} \quad \sum_{n=0}^{N-1} \left\| {\mathbf x}_n \right\|_\mathcal{A} \quad
  \subjectto \quad {\mathbf f} = {\mathbf A}_{\mathbf g}{\mathbf x}.
  \label{eq:proposed_method_ANM}
\end{equation}
Since ${\mathbf x}_n = \sum_k c_{n,k} {\mathbf a}_{n,k}$, the $n$th windowed signal ${\mathbf W}_n{\mathbf x}_n = \sum_k c_{n,k} ({\mathbf W}_n {\mathbf a}_{n,k})$ corresponds to the vertical sum of elements in Fig.~\ref{fig:dictionaries}.
The signal ${\mathbf f}$ can be reconstructed by $\sum_{n=0}^{N-1}{\mathbf W}_n{\mathbf x}_n ={\mathbf A}_{\mathbf g}{\mathbf x} $.
After solving Eq.~\eqref{eq:proposed_method_ANM}, the coefficients $c_{n,k}$ can be calculated as the line spectrum estimation using atomic norm (in Sec.~\ref{sec:lineSpectrumEstimation}).

\subsection{Algorithm for solving Eq.~\eqref{eq:proposed_method_ANM}}

We firstly reformulate it as a semidefinite programming to solve Eq.~\eqref{eq:proposed_method_ANM}.
Substituting Eq.~\eqref{eq:atomic_norm_SDP} into Eq.~\eqref{eq:proposed_method_ANM}, it can be rewritten as
\begin{align}
  \minimize_{{\mathbf x}, {\mathbf u},{\boldsymbol\nu}}& \quad \sum_{n=0}^{N-1} \frac{1}{2L}\mathrm{Tr}(T({\mathbf u}_n)) + \frac{1}{2}\nu_n \nonumber\\
  \mathop{\rm subject~to}& \quad \left[\begin{array}{cc}
    T({\mathbf u}_n) & {\mathbf x}_n \\
    {\mathbf x}_n^* & \nu_n
  \end{array}\right]
  \succeq 0 , \text{ for }n=0,\ldots,N-1 \nonumber \\
  & \quad {\mathbf f} = {\mathbf A}_{\mathbf g}{\mathbf x},
  \label{eq:proposed_method_SDP}
\end{align}
where ${\mathbf u} = \left[{\mathbf u}_0^\mathrm{T}, {\mathbf u}_1^\mathrm{T}, \ldots, {\mathbf u}_{N-1}^\mathrm{T}\right]^\mathrm{T}$ and ${\boldsymbol \nu} = \left[\nu_0, \nu_1, \ldots, \nu_{N-1}\right]^\mathrm{T}$.
We adopt the alternating direction method of multipliers (ADMM) \cite{Boyd2010} to solve Eq.~\eqref{eq:proposed_method_SDP}.
For applying ADMM to the proposed method, we introduce auxiliary variables ${\mathbf Z}_n \in \mathbb{C}^{(L+1) \times (L+1)}$ for $n=0,\ldots, N-1$ and a set corresponding to the reconstruction constraint $C = \left\{{\mathbf x} \middle| {\mathbf A}_{\mathbf g}{\mathbf x} = {\mathbf f} \right\}$.
Then, Eq.~\eqref{eq:proposed_method_SDP} is reformulated as
\begin{align}
  \minimize_{\substack{{\mathbf x} \in C, {\mathbf u},{\boldsymbol\nu} \\ {\mathbf Z}_n \succeq 0 }} & \quad\sum_{n=0}^{N-1} \frac{1}{2L}\mathrm{Tr}(T({\mathbf u}_n)) + \frac{1}{2}\nu_n \nonumber\\
  \mathop{\rm subject~to}
  & \quad\left[\begin{array}{cc}
    T({\mathbf u}_n) & {\mathbf x}_n \\
    {\mathbf x}_n^* & \nu_n
  \end{array}\right]
  = {\mathbf Z}_n, \text{ for }n=0,\ldots,N-1. \nonumber
  \label{eq:proposed_method_ADMMform}
\end{align}
The augmented Lagrangian associated with this problem is given by
\begin{align}
  {\mathcal L}({\mathbf x}, {\mathbf u}, {\boldsymbol \nu},{\mathbf Z}_n, {\mathbf \Lambda}_n) =& \sum_{n=0}^{N-1} \frac{1}{2L}\mathrm{Tr}(T({\mathbf u}_n)) + \frac{1}{2}\nu_n \nonumber \\
  &+  \left\langle {\mathbf \Lambda}_n, \left[\begin{array}{cc}
    T({\mathbf u}_n) & {\mathbf x}_n \\
    {\mathbf x}_n^* & \nu_n
  \end{array}\right] - {\mathbf Z}_n \right\rangle_\mathrm{F} \nonumber \\
  &+ \frac{\rho}{2}\left\|
  \left[\begin{array}{cc}
    T({\mathbf u}_n) & {\mathbf x}_n \\
    {\mathbf x}_n^* & \nu_n
  \end{array}\right] - {\mathbf Z}_n \right\|_\mathrm{F}^2,
\end{align}
where ${\mathbf \Lambda}_n \in \mathbb{C}^{(L+1)\times (L+1)}$ for $n=0,\ldots, N-1$ are dual variables, and $\rho > 0$ is the augmented Lagrangian parameter.
Then, ADMM consists of the following iterations:
\begin{align}
  ({\mathbf x}^{(i+1)}, &{\mathbf u}^{(i+1)}, {\boldsymbol \nu}^{(i+1)}) = \argmin_{{\mathbf x}\in C, {\mathbf u}, {\boldsymbol \nu}}\,\,{\mathcal L}({\mathbf x}, {\mathbf u}, {\boldsymbol \nu}, {\mathbf Z}_n^{(i)}, {\mathbf \Lambda}_n^{(i)}), \label{eq:X_update}\\
  {\mathbf Z}_n^{(i+1)} =& \argmin_{{\mathbf Z}_n  \succeq 0}\,\,{\mathcal L}({\mathbf x}^{(i+1)}, {\mathbf u}^{(i+1)}, {\boldsymbol \nu}^{(i+1)}, {\mathbf Z}_n, {\mathbf \Lambda}_n^{(i)}), \label{eq:Z_update}\\
  {\mathbf \Lambda}_n^{(i+1)} =&  {\mathbf \Lambda}_n^{(i)} + \rho\left( \left[\begin{array}{cc}
    T({\mathbf u}_n^{(i+1)}) & {\mathbf x}_n^{(i+1)} \\
    ({\mathbf x}_n^{(i+1)})^* & \nu_n^{(i+1)}
  \end{array}\right] - {\mathbf Z}_n^{(i+1)} \right).\label{eq:Lambda_update}
\end{align}
Introducing
\begin{equation}
  {\mathbf Z}_n = \left[\begin{array}{cc}
    {{\mathbf Z}_{\mathrm T}}_n & {{\mathbf z}_{\mathrm x}}_n \\
    {{\mathbf z}_{\mathrm x}}_n^* & {z_\nu}_n
  \end{array}\right],\quad
  {\mathbf \Lambda}_n = \left[\begin{array}{cc}
    {{\mathbf \Lambda}_{\mathrm T}}_n & {{\boldsymbol \lambda}_{\mathrm x}}_n \\
    {{\boldsymbol \lambda}_{\mathrm x}}_n^* & {\lambda_\nu}_n
  \end{array}\right], \nonumber
\end{equation}
${\mathbf z}_\mathrm{x} = \left[{{\mathbf z}_\mathrm{x}}_0^\mathrm{T}, {{\mathbf z}_\mathrm{x}}_1^\mathrm{T}, \ldots, {{\mathbf z}_\mathrm{x}}_{N-1}^\mathrm{T}\right]^\mathrm{T}$, and ${\boldsymbol \lambda}_\mathrm{x} = \left[{{\boldsymbol \lambda}_\mathrm{x}}_0^\mathrm{T}, {{\boldsymbol \lambda}_\mathrm{x}}_1^\mathrm{T}, \ldots, {{\boldsymbol \lambda}_\mathrm{x}}_{N-1}^\mathrm{T}\right]^\mathrm{T}$, Eq.~\eqref{eq:X_update} can be separately solved for ${\mathbf x}$, ${\mathbf u}$ and ${\boldsymbol \nu}$.

The proposed algorithm is summarized in Algorithm~\ref{alg:ADMM}.
The update for ${\mathbf x}$ is written as the projection onto the set $C = \left\{{\mathbf x} \middle| {\mathbf A}_{\mathbf g}{\mathbf x} = {\mathbf f} \right\}$, denoted by $P_C(\cdot)$.
It is given by
\begin{equation}
  P_C({\mathbf v}) = {\mathbf v} - {\mathbf A}_{\mathbf g}^*({\mathbf A}_{\mathbf g}{\mathbf A}_{\mathbf g}^*)^{-1}({\mathbf A}_{\mathbf g}{\mathbf v} - {\mathbf f}).
  \label{eq:projection_Affine}
\end{equation}
The computation of matrix inversion in Eq.~\eqref{eq:projection_Affine} can be avoided by the cannonical dual window $\tilde{\mathbf h}$ as
\begin{equation}
  {\mathbf A}_{\mathbf g}^*\left({\mathbf A}_{\mathbf g}{\mathbf A}_{\mathbf g}^*\right)^{-1} ={\mathbf A}_{\tilde{\mathbf h}}^*.
\end{equation}
The updates for ${\mathbf u}$, ${\boldsymbol \nu}$, ${\mathbf Z}_n$, and ${\mathbf \Lambda}_n$ can be computed in parallel for each $n$.
$T^\dagger:{\mathbb C}^{L\times L} \to {\mathbb C}^L$ in the update for ${\mathbf u}$ represents the pseudo-inverse operator of $T$,
\begin{equation}
  T^\dagger({\mathbf X})[n] = \frac{1}{2(L-n)}\sum_{k=0}^{L-n-1}\left({\mathbf X}[k, k+n] + \overline{\mathbf{X}[k+n, k]}\right).\nonumber
\end{equation}
The update for ${\mathbf Z}_n$ can be calculated by a projection onto the positive semidefinite cone ${\mathbb S}_+$, which is implemented by setting the negative eigenvalues to 0.
Since Eq.~\eqref{eq:proposed_method_SDP} is a convex optimization problem, Algorithm~\ref{alg:ADMM} can obtain the global optimal solution regardless of initialization for ${\mathbf Z}_n$ and ${\mathbf \Lambda}_n$.

\begin{algorithm}[t]
  \caption{ADMM for solving Eq.~\eqref{eq:proposed_method_ANM}}
  \label{alg:ADMM}
  \begin{algorithmic}
    \Require ${\mathbf A}$, ${\mathbf f}$, $\rho$
    \Ensure ${\mathbf x}$, ${\mathbf u}$, ${\boldsymbol \nu}$
    \State Initialize ${\mathbf Z}_n$ and ${\mathbf \Lambda}_n$ for $n=0,\ldots, N-1$
    \For{$i=0,1,\cdots$}
      \State ${\mathbf x} \leftarrow P_{C}\left( {\mathbf z}_{\mathrm x} - \frac{1}{\rho}{\boldsymbol \lambda}_{\mathrm x} \right)$
      \For{$n =1,\cdots,N$}
        \State ${\mathbf u}_n \leftarrow T^\dagger \left({{\mathbf Z}_\mathrm{T}}_n - \frac{1}{\rho}\left({{\mathbf \Lambda}_\mathrm{T}}_n + \frac{1}{2}{\mathbf I}_L\right) \right)$
        \State $\nu_n \leftarrow {z_\nu}_n - \frac{1}{\rho}\left({\lambda_\nu}_n + \frac{1}{2}\right)$
        \State ${\mathbf Z}_n \leftarrow P_{{\mathbb S}_+}\left( \left[\begin{array}{cc}
          T({\mathbf u}_n) & {\mathbf x}_n \\
          {\mathbf x}_n^* & \nu_n
        \end{array}\right] + \frac{1}{\rho}{\mathbf \Lambda}_n \right)$
        \State ${\mathbf \Lambda}_n \leftarrow {\mathbf \Lambda}_n + \rho\left( \left[\begin{array}{cc}
          T({\mathbf u}_n) & {\mathbf x}_n \\
          {\mathbf x}_n^* & \nu_n
        \end{array}\right] - {\mathbf Z}_n \right)$
      \EndFor
    \EndFor
  \end{algorithmic}
\end{algorithm}

\section{Numerical experiments}

To evaluate the performance of the proposed method, firstly, the proposed method was applied to an artificial signal that contains a sinusoid, a linear chirp, and a quadratic chirp.
The proposed method was compared to DGT with the canonical dual window, the reassignment method \cite{Auger1995}, the $\ell_1$-norm minimization \cite{Balazs2013}, and the window-wise atomic norm minimization.
The Slepian window \cite{Slepian1978} was chosen as a window ${\mathbf g}$, whose length and bandwidth were set to $2^7$ and $0.04$. The time-shifting width and the number of frequency channels were set to $a = 2^4$ and $M = 2^{10}$.
${\mathbf Z}_n$ and ${\mathbf \Lambda}_n$ in Algorithm~\ref{alg:ADMM} were initialized to the zero matrices.
Prony's method was used to estimate ${\mathbf a}_{n,k}$ from the solution of Eq.~\eqref{eq:proposed_method_SDP}.

\begin{figure*}[t]
	\centering
	\includegraphics[width=1.7\columnwidth]{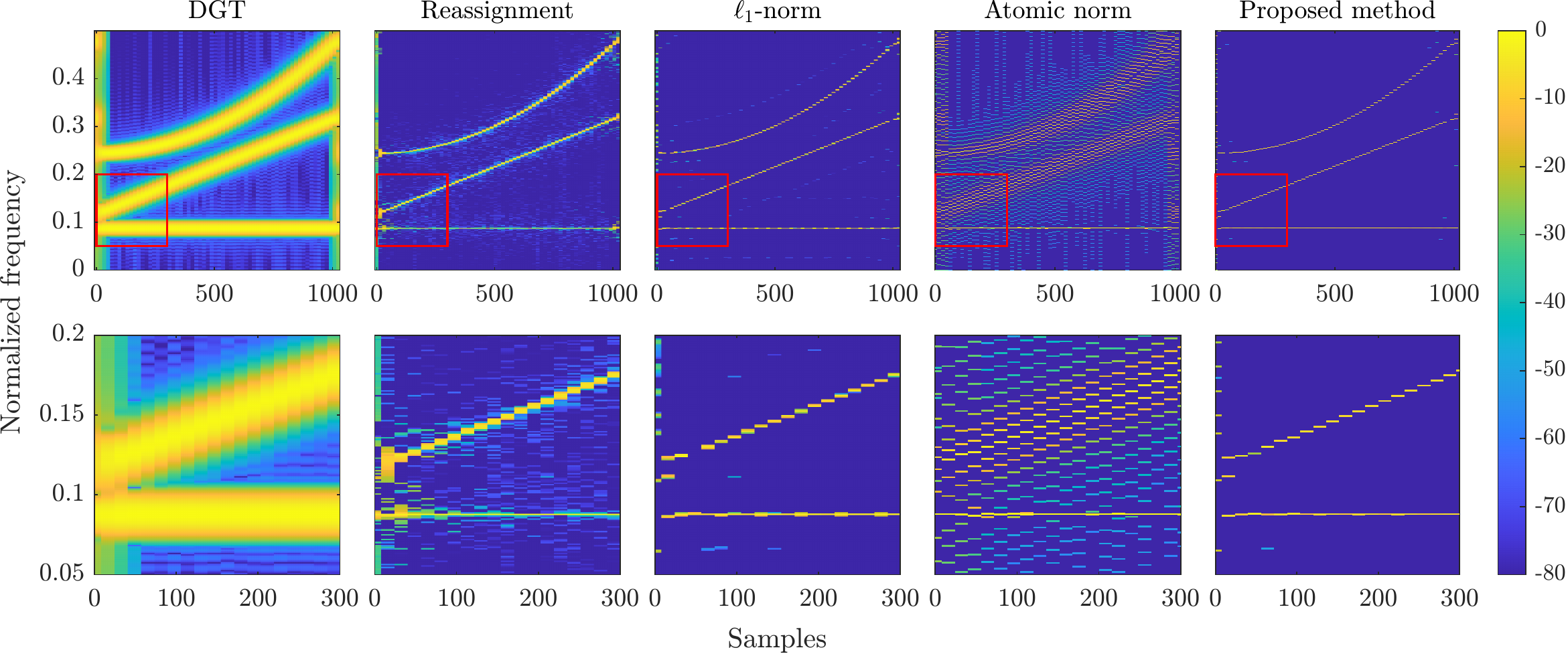}
      \vspace{-10pt}
	\caption{T-F representations of an artificial signal. Each column shows (from left to right) the T-F representations obtained by DGT, the reassignment method, the $\ell_1$-norm minimization, the window-wise atomic norm minimization, and the proposed method, respectively. The bottom row illustrates these enlargements in the red box.}
  \label{fig:Fig1}
  \vspace{-4pt}
\end{figure*}

The estimated T-F representations are shown in Fig.~\ref{fig:Fig1}.
The T-F representation obtained using the $\ell_1$-norm was better-localized than DGT and the reassignment, but it had multiple non-zero coefficients in each time index to represent a sinusoid.
The window-wise atomic norm minimization estimated a sparse representation corresponding to the sinusoid.
On the other hand, it cannot express the chirps sparsely.
The T-F representation obtained by the proposed method was the most-localized among these T-F representations.

\begin{figure}[t]
  \vspace{-1pt}
	\centering
	\includegraphics[width=0.96\columnwidth]{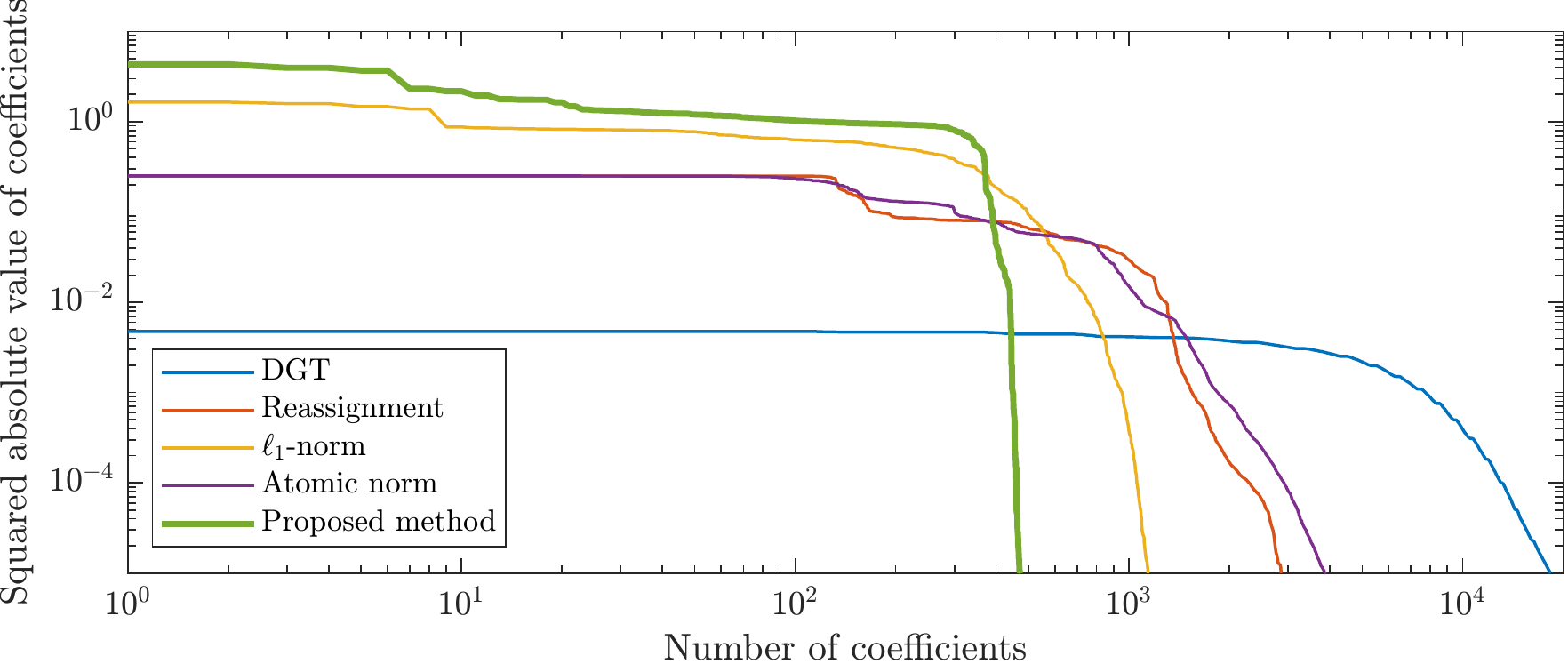}
  \vspace{-10pt}
	\caption{Squared absolute value of coefficients in decending order.}
	\label{fig:Fig2}
\end{figure}

Then, to evaluate the sparseness of these T-F representations, the squared absolute value of coefficients are plotted in Fig.~\ref{fig:Fig2}.
It can be seen that the proposed method can represent the signal using the least coefficients.
These results indicate that the proposed method provides a sparse representation using the atomic norm while taking into account the relationship among each time index.

\begin{figure*}[b]
	\centering
  \includegraphics[width=1.7\columnwidth]{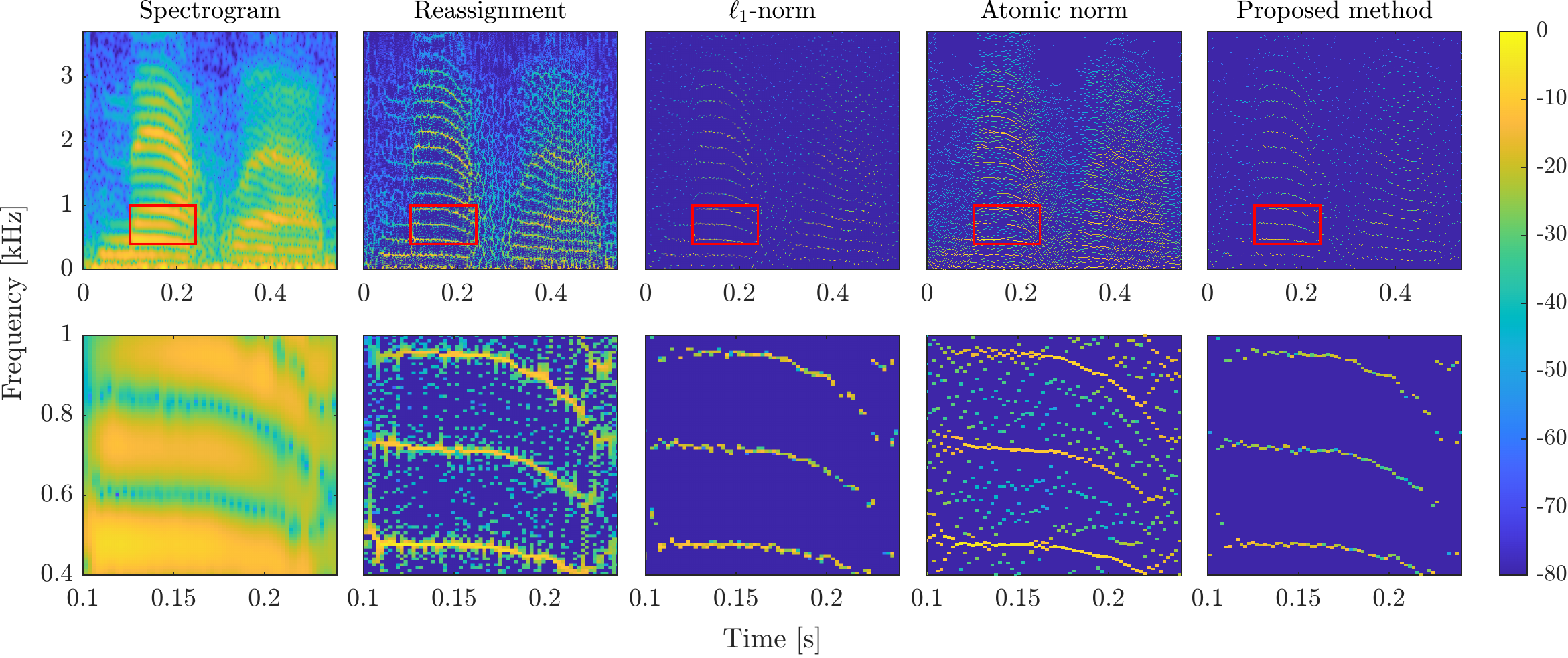}
  \vspace{-10pt}
	\caption{T-F representations of a speech signal. Each row and column represents the same as Fig.~\ref{fig:Fig1}.}
	\label{fig:spec_speech}
\end{figure*}

Finally, the proposed method was applied to a speech signal.
The settings associated with the Gabor system and the proposed algorithm were the same as in the previous experiment.
The estimated T-F representations are shown in Fig.~\ref{fig:spec_speech}.
It can be seen that the T-F representation obtained by the proposed method was the sparsest among these T-F representations.
The result suggests that the proposed method performs well also for a real audio signal.

\section{Conclusion}

In this paper, we proposed the method of estimating sparse T-F representation via the atomic norm minimization.
The proposed method estimates a sparse T-F representation without discretization of frequency using the atomic norm.
The experimental results show that the proposed method can estimate a sparser T-F representation than the existing methods.
Future work includes applications of the proposed method to denoising and mode decomposition.

\bibliographystyle{IEEEbib_AuthorAbbreviation}
\bibliography{ICASSP2021}

\end{document}